\documentclass[12pt,a4paper]{article}
\usepackage[T2A]{fontenc}
\usepackage{amsfonts}
\usepackage{amsmath}
\usepackage{amssymb}
\usepackage[english]{babel}
\usepackage{graphicx}
\begin{document}
\title{Unfaithful embedding of causal sets}
\author{Alexey L. Krugly\thanks{Quantum Information Laboratory, Institute of Physics and Physical Technologies, Moscow, Russia; akrugly@mail.ru.}}
\date{} \maketitle
\begin{abstract}
In the causal set approach, spacetime is a continuous approximation of a faithful embedded causal set. But the faithful embedded causal set describes the empty spacetime and cannot describe matter. Only faithful embedding of coarse grained causal set is necessary for the emergence of spacetime. The initial causal set can be unfaithfully embedded. The following properties of the universe can be the consequences of the unfaithful embedding: antiparticles, fields of interactions, the inhomogeneous distribution of the matter.
\end{abstract}
The causality among events is a most fundamental structure of spacetime, and causal sets have been proposed as discrete models of quantum gravity in the Causal Set Programme \cite{Myrheim,'t Hooft,Sorkin2005,Dowker2006,Henson2009,Wallden2010}. A causal set is a locally finite partially ordered set, that is, a set of events with a partial order relation `$\preceq$' which is:
\begin{itemize}
\item reflexive: for any event $a$, $a\preceq a$;
\item anti-symmetric: if $a\preceq b$ and $b\preceq a$ then $a=b$;
\item transitive: if $a\preceq b$ and $b\preceq c$ then $a\preceq c$;
\item locally finite: the number of events between any two events is finite.
\end{itemize}

One of the goals of the Causal Set Programme is to investigate the emergence of continuous spacetime as an approximation of some kind of causal sets. The main idea is a faithful embedding of causal sets \cite{BMSorkin}. We should seek a manifold $M$ (with a time-oriented Lorentzian metric $g_{ij}$ free of closed timelike or null curves), and an embedding $f:C\longrightarrow M$ of the causal set into the manifold. Let us call an embedding `faithful' if the following conditions are satisfied:
\begin{itemize}
\item the causal relations induced by the embedding agree with those of $C$ itself, i.e., $f(a)\in J^- (f(b))$ iff $a\prec b$, where $J^-(c)$ denotes the set of points of $M$ to the causal past of $c$;
\item the embedded points are distributed uniformly with unit density;
\item the characteristic length over which the continuous geometry varies appreciably is everywhere much greater than the mean spacing between embedded points.
\end{itemize}

The faithful embedding is related with the concept of 'sprinkling', which refers to the process of creating a causal set $C(M)$ by generating points in a manifold $M$ with Lorentzian metric according to a Poisson distribution of unit density. The discrete structure inherit causality from the continuous one, and the nature of the distribution guarantees that the number of points found in a region of $M$ is proportional to the volume.

This concept justifies the equation `geometry = order + number'. But there is the problem with the description of the matter. The faithfully embedded causal set possesses a very weak structure. It approximates empty spacetime and cannot describe the matter. It is necessary to put the matter to this causal set ad hoc.

The conformal structure of continuous spacetime is derived from the topology of continuous timelike curves \cite{Hawking 1976,Malament 1977}. We must have continuum of timelike curves in each spacetime point. Causal sets can be drawn as directed, acyclic graphs. If we describe the elements of the faithfully embedded causal set as the vertices of the graph, and segments of the timelike curves as edges these vertices must have an infinite valency. There is no way to associate the finite-valency graph to the causal set consistently with Lorentz invariance \cite{gr-qc/0605006}.

Let us consider the alternative point of view. It based on the relational concept of spacetime. Only the matter exists and usual spacetime relations should be derived from interactions between elementary particles when we consider large systems of the elementary particles. There is no empty spacetime between matter, no spacetime relations without interactions. In this case, the causal set must describe the matter and its elements are material objects. The physical meaning of the timelike curve is a world line of some particle. If there are infinitely many world lines in the spacetime point there are infinitely many particles in this point. Consequently the spacetime point corresponds to some subset of the causal set. This subset must contain many elements of the causal set. Similarly, in thermodynamics the infinitesimal volume of a gas contains many molecules.

Spacetime can be emerging by the following steps. The causal set forms the particles as some repetitive structures (see e.g. \cite{1004.3128,1004.5077}). These structures are approximated by the world lines. Some subsets of the causal set can be identified as spacetime points only if these subsets are crossed by many world lines (approximately infinitely many world lines).

We can consider these subsets as elements of a new causal set or vertices of a new graph. This is the coarse graining of the initial causal set (the initial graph). The faithful embedding of the initial graph is not useful. Its vertices can have the low valency. The vertex of the coarse grained graph includes many vertices of the initial graph and has the high valency. We can consider the sequence of the coarse graining such that the valency of vertices increases. There is Lorentz invariance only in the limit of the infinite valency. The vertex of the finite valency corresponds to the finite number of particles. Such system has a preferred reference frame. This is center-of-mass frame. We must have the faithful embedding only in the limit of the coarse graining.

The unfaithful embedding of the initial graph can have observable properties.

The matter is distributed in the universe with a various density. This is the first observable property. Two equal volumes can contain different numbers of particles. Consequently, the number of causal set elements does not correspond to the spacetime volume. The spacetime metric is determined by chronometric measurements \cite{Synge 1960}. All spacetime measurements are reduced to measurements of intervals of proper time. If the particle is some repetitive structure the proper time is the number of cycles of this structure timed by some coefficient. We can choose one kind of particles as a standard clock with an arbitrary coefficient. If we determine the metrical structure in some spacetime volume by the standard clocks we can determine these coefficients for any particles in this volume and can use these particles as the standard clocks in subsequent measurements.

The second observable property is antiparticles. The causal structure of spacetime must correspond to the causal order of the coarse grained graph. But some pairs of vertices of the initial graph can have inverse causal order to the macroscopic time. We have some statistical procedure to determine the causal order of the coarse grained graph. Let us consider the simple example of such procedure. Two vertices $v_1$ and $v_2$ of the coarse grained graph consist of the set $\{a_i\}$ and $\{b_j\}$ of the elements of the initial causal set, respectively. Let us consider the set of all pairs $A=\{a_i b_j\}$. The cardinality of $A$ is denoted by $N$. In the general case, we have three subsets of this set: $A_1=\{a_i b_j|a_i\prec b_j\}$, $A_2=\{a_i b_j|a_i\succ b_j\}$, and $A_3=\{a_i b_j|a_i\| b_j\}$ (a causally unrelated pair of elements is denoted by $a_i\| b_j$). The cardinalities of these subsets are denoted by $N_1$, $N_2$, and $N_3$, respectively. By definition, $v_1\prec v_2$ iff $N_1\sim N$, $N_1\gg N_2$, and $N_1\gg N_3$. If $N_2\ne 0$ there are the pairs of causally related elements of the initial graph such that their causal order is opposite to the causal order of spacetime points. In this model the number of antiparticles is less than the number of particles in the universe.

The third observable property is the fields of interactions. The idea corresponds to the concept of a direct interparticle action, which is alternative to the conventional field theory \cite{Tetrode 1922, Fokker 1929, WF 1945, WF 1949}). This list of references is by no means complete. In such approach, the concept of a field is excluded from the primary concepts because it has meaning only in the presence of the classical spacetime. Without the last, it is possible to use only direct relations between interacting particles.

\begin{figure}
	\centering	
		\includegraphics[trim=8cm 16cm 8cm 7cm]{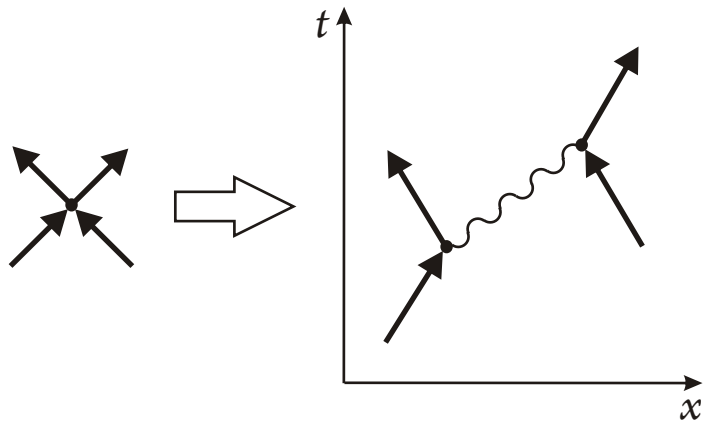}
	\caption{The tetravalence vertex of the directed, acyclic graph corresponds to two trivalent vertices after the unfaithful embedding to continuous spacetime. Spacetime is denoted by two coordinate axises. The connection of two trivalent vertices is denoted by a wavy line.}
	\label{fig:fig1}
\end{figure}

Suggest the one vertex of the initial graph can be included in several vertices of the coarse grained graph. Consequently this element corresponds to different spacetimes points. Let us consider the simple example (Fig.\ \ref{fig:fig1}).
Let the direct interaction of two particles be the vertex of the graph. The valency of this vertex is equal to 4. Let this vertex be included in two different vertices of the coarse grained graph and correspond to two spacetime points. In this case we must describe the connection between this points as a quantum of interaction. There are two trivalent verices instead one tetravalence vertex. In this example, the quantum of interaction is the consequence of the unfaithful embedding. The vertices can possess ambiguous spacetimes coordinates. Consequently we must integrate with respect to these coordinates.

Probably the most fruitful way in the Causal Set Programme is the investigation of the directed, acyclic graphs such that the vertices possess low valency. The local structure of this graph must correspond to the individual particles and their interactions. These particles must dynamically arise as self-organized structures. Spacetime describes the relations between the particles averaged over much number of the particles. Spacetime geometry must be derived from an underlying dynamics.

\end{document}